# ContractTrace: Retracing Smart Contract Versions for Security Analyses

Fatou Ndiaye Mbodji[1], Boladji Vinny Adjibi[2], Moustapha Awwalou Diouf[1],
Gervais Mendy[3], Kui Liu[4], Jacques Klein[1], Tegawendé F. Bissyandé[1]
[1]SnT – University of Luxembourg, [2]Georgia Institute of Technology, [3]Université Cheikh Anta Diop, [4]Huawei

***Abstract*—Due to the inherent immutability of blockchain technology, smart contract updates require their deployment at new addresses rather than modifying existing ones, thus fragmenting version histories and creating critical blind spots for analyses. Indeed, for example, this fragmentation severely hinders security researchers' ability to track vulnerability lifecycles across contract versions. While platforms like Etherscan provide detailed information about Ethereum smart contracts, they lack crucial functionality to trace predecessor-successor relationships within smart contract lineages, preventing systematic analysis of how vulnerabilities emerge, propagate, and potentially remain unresolved across versions.**

**To address the challenge of tracing smart contract lineages, we adopt a Design Science Research (DSR) approach and introduce `ContractTrace`, an automated infrastructure that accurately identifies and links versions of smart contracts into coherent lineages. This tool enables the construction of *lineageSet*, an up-to-date, open-source dataset specifically designed to support security research on vulnerability, defect or any other property's evolution patterns in smart contracts.**

**Through a secuity-focused case study we demonstrate how `ContractTrace` reveals previously obscured vulnerability life-cycles within smart contract lineages, tracking whether critical security flaws persist or get resolved across versions. This capability is essential for understanding vulnerability propagation patterns and evaluating the effectiveness of security patches in blockchain environments. In the evaluation phase of our DSR approach, we validated our lineage detection methodology against an alternative approach using Locality-Sensitive Hashing (LSH) to cluster contract versions, confirming the security relevance and accuracy of our technique.**

## I. Introduction

Smart contracts, self-executing programs deployed on blockchains, have gained traction across industries like finance, healthcare, and real estate, where trust, security, and reliability are paramount. However, the inherent immutability of blockchain technology presents considerable challenges when tracking the evolution of these contracts. Therefore, updates to smart contracts require deployment to new addresses, effectively severing the links between different versions. This complicates the tracing of contract lineages as a sequence of versions of the same smart contract. In the literature, authors leveraging lineages for smart contract analyses claim to build on similarity metrics for building such lineages [1], [2]. We postulate that, due to the heavy reuse of code across smart contracts [3], illustrated by the high rates of code duplication within the Ethereum ecosystem, as well as the overall similarity of smart contract behaviors, similarity-based approaches will lead to unreliable lineages. Unfortunately, datasets described in the literature are not shared with the community for assessment or even reuse, hindering broad research on smart contract evolution.

To address the literature gap on smart contract lineages, we propose to build a large-scale, extensible, and open dataset of smart contracts where lineages are tracked. To that end, we rely on a conservative approach based on the concept of *proxy* in smart contract deployment. Proxy contracts, which act as intermediaries and redirect users to the latest contract version, offer a solution to the immutability problem in terms of interaction with the smart contract new versions. Unfortunately, even with proxies, establishing lineage across different contract addresses remains difficult. Existing platforms of smart contracts corpora, like Etherscan and smart-corpus [4] provide detailed information about deployed contracts but do not explicitly trace predecessor-successor relationships.

In this paper, we introduce `ContractTrace`, an infrastructure designed to identify and collect smart contract lineages systematically. `ContractTrace` leverages proxy contracts to trace contract updates accurately and produces *lineageSet*, a comprehensive and open-source dataset of smart contract lineages. This dataset facilitates large-scale research on contract evolution and provides new insights into how smart contracts are maintained and updated over time. Currently, *lineageSet* contains 1 055 smart contracts distributed across 347 lineages. It is openly available on Github:

https://anonymous.4open.science/r/sclineages-A9A2

*Community Benefits.* Our infrastructure serves as a valuable resource for software engineering and security research, offering a comprehensive dataset of linked smart contract versions that can be leveraged for various analytical purposes. This enables researchers to explore contract evolution, vulnerability management, and the reliability of existing lineage construction techniques.

- **Case study #:** *Identifying vulnerability life-cycle for smart contracts.* Vulnerability fixes in software are often silent. Since *lineageSet* includes lineages of production smart contracts, we apply vulnerability detection tools on the different versions of the lineages to retrieve the code changes that have led to vulnerability warning apparitions and disappearances.
- **Evaluation:** *Revisiting the reliability of similarity-based construction of smart contracts*. Given the conservative

way *lineageSet* was built, we can consider it as ground truth for validating approaches for lineage construction. we assess the reliability of similarity computation as proposed in prior literature. We consider the Locality-Sensitive Hashing (LSH) method implemented in the Etherscan search engine as a similarity computation model. This evaluation is for validating the relevance of our methodology.

Overall, the main contributions of our work are as follows:

1) `ContractTrace`: we propose a straightforward but novel approach for building lineages, leveraging proxy contracts to ensure accurate lineage tracking, and overcoming the challenges associated with blockchain immutability.
2) lineageSet: we present *lineageSet*, a comprehensive and open-source dataset of smart contract lineages. This rich dataset stands out in the literature due to its:
   a) *Extensive Scope*: Encompassing 1,055 smart contracts *lineageSet* provides a vast landscape for research exploration.
   b) *Open Accessibility*: In contrast with prior work that has built lineages, *lineageSet* is freely available on GitHub and will foster reproduction studies and collaboration to advance research on smart contract analysis.
   c) *Ground-Truth Foundation*: Due to its conservative construction method, *lineageSet* serves as a reliable benchmark for validating future lineage construction approaches.
   d) *Security analysis*: *lineageSet* serves to examine the vulnerability life-cycle in smart contracts, allowing for a detailed analysis of how vulnerabilities emerge and are resolved across contract versions.

In the following section, we define and present key concepts used throughout the paper. We also describe our research method, Design Science Research (DSR), which is applied in the subsequent sections, followed by the discussion, related works, and finally, the conclusion.

## II. Background and Foundation

### A. Ethereum Smart Contract

In 1997, Szabo envisioned smart contracts as self-executing programs that automate agreements between parties, eliminating the need for a trusted third party [5]. The emergence of blockchain technology in 2008, as documented in the white paper of Nakamoto [6], provided a robust platform for the implementation of smart contracts. Blockchain technology facilitates disintermediation by eliminating the need for third-party intermediaries in transactions, fostering trust among participants in a decentralized network. At its core, a blockchain is a chronologically ordered sequence of data blocks, referred to as a distributed ledger. This ledger is managed collaboratively by a peer-to-peer (P2P) network, ensuring decentralization and eliminating the need for participants to trust each other. The integrity of the data within the blockchain is cryptographically secured, with each block linked to the preceding one using a cryptographic hash, thus ensuring immutability and tamper-proof data storage.

With the advent of blockchain technology, a smart contract, as formally described by Nick Szabo [5], has been effectively and fully implemented. Smart contracts are defined as software programs that, once deployed on a blockchain platform, automate the execution of agreement terms between parties, thereby eliminating the need for a third party.

Blockchain technology is currently implemented in various platforms, our study specifically focuses on Ethereum [7], the most prominent platform known for its extensive use of publicly-operated smart contracts. Ethereum offers a virtual machine capable of executing smart contracts utilizing blockchain technology.

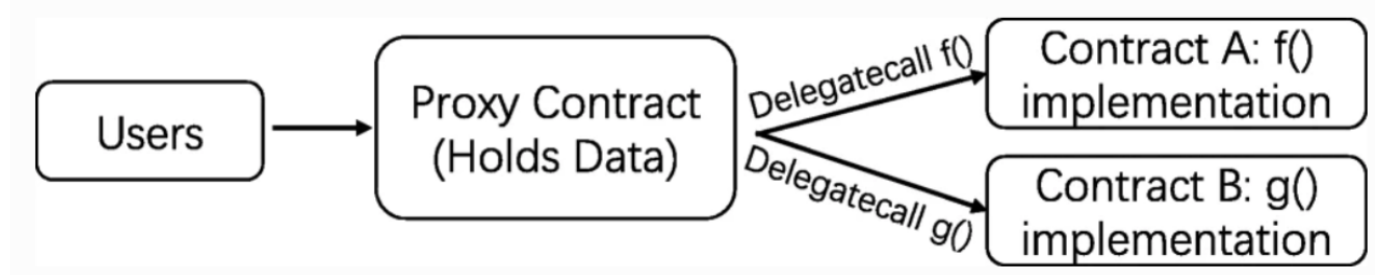

Fig. 1. An Example of the Upgradable Contract from Chen et al. [8].

### B. Proxy pattern: A Nuance Between Immutability and Upgradable Smart Contracts

Smart contracts inherit the key characteristics of blockchains, such as decentralization and immutability. Once a smart contract is built and deployed at a specific address, it cannot be modified. Any new version of a contract must be redeployed at a new address.

To provide an interface for accessing updated code without the need for the new address, a proxy pattern can be implemented. When this pattern is applied, users interact with the proxy contract, which cannot be modified. However, the proxy is able to point to a configurable address, allowing it to call another contract known as the callee contract. Thus, the parts of the code that are subject to change are located in the callee contract. When an upgrade is needed, the developer redeploys a new version of the callee contract with a new address, and the proxy contract reconfigures its callee contract address accordingly. Typically, an admin user has the privilege to perform this address configuration. Therefore, users interact with the same proxy contract even when upgrades occur.

Fig. 1 illustrates this method, which is facilitated by constructs proposed by smart contract programming languages. For instance, in Solidity, the `delegatecall` allows a contract caller (or proxy) to use the code of a callee contract while remaining in the global context of the caller.

Our work builds upon this proxy concept to track smart contract updates.

### C. Smart contracts versions

*Predecessor/Successor Relationships in Smart Contracts:* We adopt the notion of predecessor/successor from prior work [1], where a predecessor is defined as the most recent version preceding the successor.
*Smart Contract Lineage:* In this study, lineage refers to a set of smart contracts where each contract can be paired with another based on a predecessor/successor relationship.

### D. Computing Smart Contract Similarity:

Several reasons contribute to code cloning in deployed smart contracts, such as deploying new versions, the open source characteristic of many smart contracts, the simplicity of copying code fragments instead of writing them from scratch, etc. [3]. The literature contains numerous studies on Ethereum smart contract code clone detection, including clone detection techniques such as SmartEmbed [9], Deckard [10], Nicad [11], and LSH-based approaches [12]. The LSH-based method is implemented in Etherscan, which fingerprints smart contracts and computes their similarity levels, categorizing them as low, medium, or high.

### E. Dynamic and Up-to-Date Repositories for Smart Contract Analysis in Empirical Software Engineering

Etherscan[1] is a widely used Ethereum block explorer that allows developers to submit the source code of their smart contracts, making this code available for verification and transparency. This platform provides a comprehensive database essential for developers to verify and analyze smart contracts deployed on the Ethereum blockchain.
Smart Corpus [4] aims to be an organized and up-to-date repository where developers can systematically access Solidity source code and other metadata about Ethereum smart contracts. This repository facilitates the retrieval of detailed information on smart contracts and their software metrics, streamlining the process for developers and researchers.
Tools like Etherscan and Smart Corpus, which provide accessible and current information on smart contracts, are crucial for advancing empirical software engineering research. They enable a thorough understanding of real contracts deployed on Ethereum, supporting the development of more robust and transparent smart contracts. These repositories offer a variety of information about smart contracts but do not provide the lineage of contracts. Our study is situated in the same context and aims to contribute information about smart contract lineages.

[1] https://etherscan.io/

Hence, our objective is to create a repository where access to versions of smart contracts deployed on Ethereum is streamlined.

## III. Design Science Research (DSR) as Method

We follow the Design Science Research (DSR) approach. The DSR process includes six steps: problem identification and motivation, objectives for a solution, design and development, demonstration, evaluation, and communication. These are the essential activities that make up the DSR process:

a) **Problem Identification and Motivation**: This phase involves defining the research problem and justify the importance of solving it. We recognized the challenge of tracking smart contract lineages. As a specific example, this gap hampers the ability to study how vulnerabilities emerge and evolve over time. This step is developed in detail in (IV-A) and (IV-B).
b) **Define the Objectives for a Solution**: The goal of this step is to formulate solution objectives based on the problem and feasible solutions to define the desired outcome. In this study, we aimed to build an infrastructure to track and link contract versions, enabling research on their evolution and security. Further details of this step are found in (IV-C)
c) **Design and Development**: This step is to create the artifact by defining its desired functionality and architecture. We built `ContractTrace` to track smart contract lineages using proxy contracts and created *lineageSet*, a dataset for research purposes. This is covered in (V)
d) **Demonstration**: This step is for utilizing the artifact to address real-world instances of the problem. We applied `ContractTrace` in a case study on the lifecycle of vulnerabilities in smart contracts. This is demonstrated in (VI).
e) **Evaluation**: The evaluation tests the effectiveness of the artifact. We validated our approach by applying Locality-Sensitive Hashing (LSH) for clustering contract versions, confirming the approach robustness. Details of this evaluation are in (VII)
f) **Communication**: The final step involves sharing the findings and the artifacts design to relevant stakeholders. In this paper, we consider the communication step as the dissemination of results through publication of this paper.

## IV. Problem, Motivation, and Objectives

This section presents the application of the first two phases of the Design Science Research approach.

### A. Studies Highlighting the Need for Contract Lineage Information

Our literature review identified several studies that propose algorithms leveraging code similarity metrics for the classification of smart contract versions.

*Chen* et al. *[1].* The predecessor/successor concept was used in 2021 by Chen *et al.* to group smart contracts into pairs of predecessor/successor. To obtain predecessor/successor pairs for their research question, which was ”Why do smart contracts self-destruct?”, they consider that the predecessor had to have executed the self-destruct function and that the successor must be deployed by the same creator address as its predecessor, and it must also have similar functionality to that of the predecessor. Hence, to determine the similarity rate between two contracts, they employed the SmartEmbed tool [9] and chose 0.6 as the minimum value for the similarity rate between the predecessor and successor. To filter out false pairs, they performed a manual check. Their work presents the following limitations: (1) Reproducing their work is difficult due to manual verification. (2) Only contract versions that executed a self-destruct function are included in their shared dataset. (3) Another limitation of their approach is the risk of false predecessor/successor pairs due to the manual verification process.

*Huang* et al. *[2].* To test their technique, which involves guiding smart contract updates by detecting code smells, they needed lineages of smart contracts. To construct a dataset of lineages, they used two criteria: the submission of contracts by the same creator and a similarity degree of 0.7 or higher between the contracts. Compared to Chen *et al.* , their approach requires a higher similarity rate. However, given the strong tendencies for code copying, using similarity to determine contract lineages may include contracts that do not actually belong to the lineage.

### B. Motivation for Leveraging the Proxy Pattern for Lineage Construction

Our study is motivated by the need to establish up-to-date repositories that facilitate easy and rapid access to versions of smart contracts. These repositories can serve as valuable resources for analyses in studies such as those conducted by Chen *et al.* [1] and Huang *et al.* [2]. This dynamic approach aims to streamline the process of accessing and analyzing smart contracts deployed on Ethereum, thereby supporting empirical software engineering research.

To the best of our knowledge, there is no existing infrastructure that provides a comprehensive, openly accessible, and systematically organized dataset of smart contract lineages. Our main objective is to establish an infrastructure for building a publicly available dataset of smart contracts meticulously grouped and ordered within their respective lineages. To ensure the integrity of the research process and facilitate further exploration within the smart contract analysis community, the dataset will be both reproducible and reliable.

To address this gap, we propose an approach using the Proxy Pattern. Given that contract calls through proxies are trackable, we plan to leverage this feature to develop a method for constructing lineages. This approach will enable the creation of a comprehensive and systematically organized repository, enhancing the accessibility and utility of smart contract data for researchers and developers alike.

### C. Objectives

Our study aims to create a comprehensive and systematically organized repository of smart contracts that will be publicly available and up-to-date. By leveraging the Proxy Pattern, we anticipate several key outcomes:

- Enhanced accessibility, allowing developers and researchers to benefit from easy and rapid access to a well-organized dataset of smart contracts, which will facilitate empirical software engineering research.
- Detailed lineage information, providing meticulously grouped and ordered smart contracts within their respective lineages, offering valuable insights into the evolution and relationships of smart contracts.
- Reproducibility and reliability, ensuring the dataset supports the integrity of the research process and enables further exploration within the smart contract analysis community.
- Supporting empirical study: By providing comprehensive data on smart contracts deployed on Ethereum, the repository will serve as a valuable resource for conducting empirical studies.

Additionally, we will test the applicability of this repository in analysis scenario to demonstrate its practical utility and effectiveness in real-world research contexts. These results will contribute significantly to the field of smart contract analysis, supporting the development of more robust and transparent smart contracts.

## V. `ContractTrace`: Design and Development

This section presents the protocol for collecting smart contract lineages, its implementation, and the resulting infrastructure.

### A. Experimental setup

Here, we present the collection of smart contracts and the rules defined to sort them into groups according to their belonging to the same lineage.

*Step 1: Data Collection:* The infrastructure relies on Etherscan to collect smart contracts. We target the contracts that are called via *proxy* contracts.

**Targeted contracts:** In our collection, we focus only on contracts updated by using the proxy technique.

This is because, unlike other upgrade methods, proxies inherently track interactions with the contracts they govern. Hence, this design choice is to reduce the likelihood of false positives when classifying contracts within lineages. Unfortunately, for contracts updated without being associated with a proxy, we have not found a common denominator to retrieve a link between the various versions of the same contract. Prior works retrieved versions of contracts regardless of the technique used to update them. Still, their approaches require manual investigations that can be error-prone (e.g., in [1]) or rely on design choices leading to small datasets (e.g., in [2]). In our study, we aim to propose an approach that yields a large, evolving, and reliable set of contract versions. Consequently, automation is key. Hence, we only target contracts called in proxy contracts.

**Data sources: Etherscan and BigQuery** While Etherscan offers access to details of smart contracts, including source code and transaction information, it only provides direct access to a limited set of the most recently verified contracts (500 latest verified contracts [2].). Verified contracts are contracts for which Etherscan has checked that their provided source code matches with their bytecode deployed on the blockchain at the given address. Since our goal is to find lots of proxy contracts, we do not rely solely on Etherscan. To overcome this limitation, we combine Etherscan with the publicly available Ethereum smart contract dataset on Google BigQuery [3], which offers a broader range of contract addresses.

**Collecting proxies and their callee contracts:** The proxy paradigm allows us to see the executing contract that receives the calls throughout the lifetime of the contract. Our approach leverages the fact that, the proxy calls the 'upgradeProxy' method of the interfacing contract, with the address of the new contract. We use Google BigQuery to identify all the contracts that called that function. The function has a specific keccak-256 value that could be computed to represent the signature of the function. Once the methods are identified, we automatically collect the addresses. These addresses are then used to query Etherscan and collect the details (in particular, the source code) of the callee smart contracts. Upon getting that list of contracts, we proceed to classify them into their respective lineages based on a defined set of criteria, which will be explained in the next step.

*Step 2: Design choice for lineages formation and Contract versioning:* After collecting smart contracts that were called by proxy contracts, we classified them into lineages. In the following, we describe the classification process.

*Notation Key:* In terms of notations, we have:

- $C$ is the set of contracts accessible in the data source (i.e., available in Etherscan);
- $\Sigma$ is a contract lineage, defined as $\Sigma = S_1 S_2 ..... S_n$, where the contract $S_i \in C$ is the $ith$ versions of the contract code called by a unique contract proxy $P$;
- P is indeed unique, and $proxy(\Sigma) = P$; i.e., each lineage lies with a unique proxy contract.
- $P$ is a proxy means $P$ is a contract which executes a delegatecall instruction so, $instructionsE(P) = I_1 I_2 ..... I_n$ where instructions $I_i$ are the executed instructions in $P$'s source code. Hence,$\exists I_x \in instructionsE(P)$ such as $I_x$ contains a delegatecall instructions.
- We also have $address(S)$ which gives the address of a smart contract $S$;
- We note by $len(\Sigma)$ the number of contract versions in the lineage $\Sigma$.

*Classification Rules:*
A contract $S$ is classified within a lineage $\Sigma$ based on the following set of rules:

- **Rule 1: Lineage Members are Callees of the Lineage's Proxy**
  $(\Sigma = (S_1 S_2 ..... S_n)$ and $proxy(\Sigma) = P) \implies \forall S_i \in \Sigma, \exists I_x \in instructionsE(P)$ such as $I_x$ contains a delegatecall instructions and the callee address in an execution of $I_x$ is $address(S_i)$. So, each lineage relies on one proxy and in a lineage, we cannot find a contract that was not called by the proxy.
- **Rule 2: A lineage has at least 2 versions of contract** : $len(\Sigma) \geq 2$
- **Rule 3: The versions are in chronological order and there is no overlap in the activity period of the lineage versions**
  We have: $firstDelegateCall(P, S)$ (respectively $lastDelegateCall(P, S)$ ) gives the date of the first (respectively last) execution of a delegatecall instruction in the proxy contract $P$ where the callee address in the delegatecall is $address(S)$;
  $(\Sigma = S_1 S_2 ..... S_n$ and $proxy(\Sigma) = P) \implies \forall S_i \in \Sigma, \ lastDelegateCall(P, S_i) < firstDelegateCall(P, S_{i+1})$ with $i < len(\Sigma)$. The updated version should replace the previous version then, it starts its activity after the previous stop its activity. Our goal in doing this work was to end up with a dataset that is of high confidence and can be trusted for many various tasks. The linear model that we decided to follow and the rules that we set make sure that we can confidently affirm that the said contracts are related. A tree-like structure would imply a lot of dependencies that could be hard to verify that they are related.
- **Predecessor/successor pairs of contracts:** These rules are applied to classify contracts in their lineage. Then, in each lineage, we classify contracts in couples of predecessor/ successor. A predeces-sor (respectively a successor) of a contract is the contract corresponding to the most (respectively the least) recent version which precedes (respectively succeeds) the contract. $\Sigma = S_1S_2.....S_n \implies \forall S_i \in \Sigma, predecessor(S_i) = S_{i-1}$ with $i > 1$ and, $successor(S_i) = S_{i+1}$ with $i < len(\Sigma)$.

[2] https://etherscan.io/contractsVerified/
[3] https://cloud.google.com/blog/products/data-analytics/ethereum-bigquery-public-dataset-smart-contract-analytics?hl=en

**Predecessor/successor pairs of file.sol:** Each contract version in a lineage is identified by one address; some contracts have more than one file with a ".sol" extension. The ".sol" extension refers to Solidity code files. Indeed, a contract $S = f_1f_2.....f_n$ where $f_i$ is a file with extension ".sol".

We also classify files from pairs of predecessor and successor contracts to facilitate analyses based on code changes. Each file is identified by its filename, which may change from one version to another. However, when examining predecessor-successor contract pairs, we observed that filename changes within the same subdirectories tend to involve only a small number of characters. For example, we found files named "LandRegistryV2.sol" and "LandRegistryV3.sol" accessible via the same subdirectories in two consecutive versions of smart contracts. We account for the possibility that these minor changes can appear in the filename, considering a similarity threshold with a two-character difference. We implement this proposal to build our infrastructure and collect smart contracts lineages.

### B. *Resulting dataset:* lineageSet

Following the implementation and execution of the two-step process outlined above, we obtain *lineageSet*. This dataset provides details on smart contract lineages built based on proxy contracts and their called contracts. *lineageSet* aims to contribute to the field of smart contract software engineering research. Not only will *lineageSet* provide much-needed data to the community, but also it ensures that competing approaches are benchmarked transparently on the same diverse and large-scale data. *lineageSet* will keep growing because the collection is conducted regularly using Etherscan and BigQuery to collect real-world contracts deployed on the Ethereum platform. The implementation and results are open access[4]. We encourage users of our dataset to share their analysis results with the community by adding their links to the dedicated page in the repository. This page aims to foster collaboration between researchers on smart contract software engineering, promote open data and up-to-date datasets, and enable comprehensive analyses.

**Figures:** After our first execution, we have figures reported in I

Our dataset *lineageSet* had 1055 smart contract addresses which were called in 347 distinct proxy addresses. Then, we identified 347 lineages having 706 predecessors/successors pair. Only 48.48% of smart contracts are open source. The total number of Solidity files in these contracts is 6049.

[4]https://anonymous.4open.science/r/sclineages-A9A2

TABLE I
SUMMARY OF DATASET FIGURES

| Metric | Value |
|---|---|
| Lineages identified | 347 |
| Pairs of predecessor/successor contracts | 706 |
| All smart contracts | 1055 |
| Percentage of open source smart contracts | 48.48% |
| Solidity files in open source smart contracts | 6049 |
| Percentage of updated files | 17.79% |
| Pairs of of predecessor/successor files | 3450 |
| Average days to deploy new version | 23 |
| Files in predecessor/successor files pairs | 88.68% |
| Average similarity rate between files paired | 98% |
| Files pairs with similarity rate $\geq$ 90% | 97% |
| Number of functions classified in pair | 43964 |

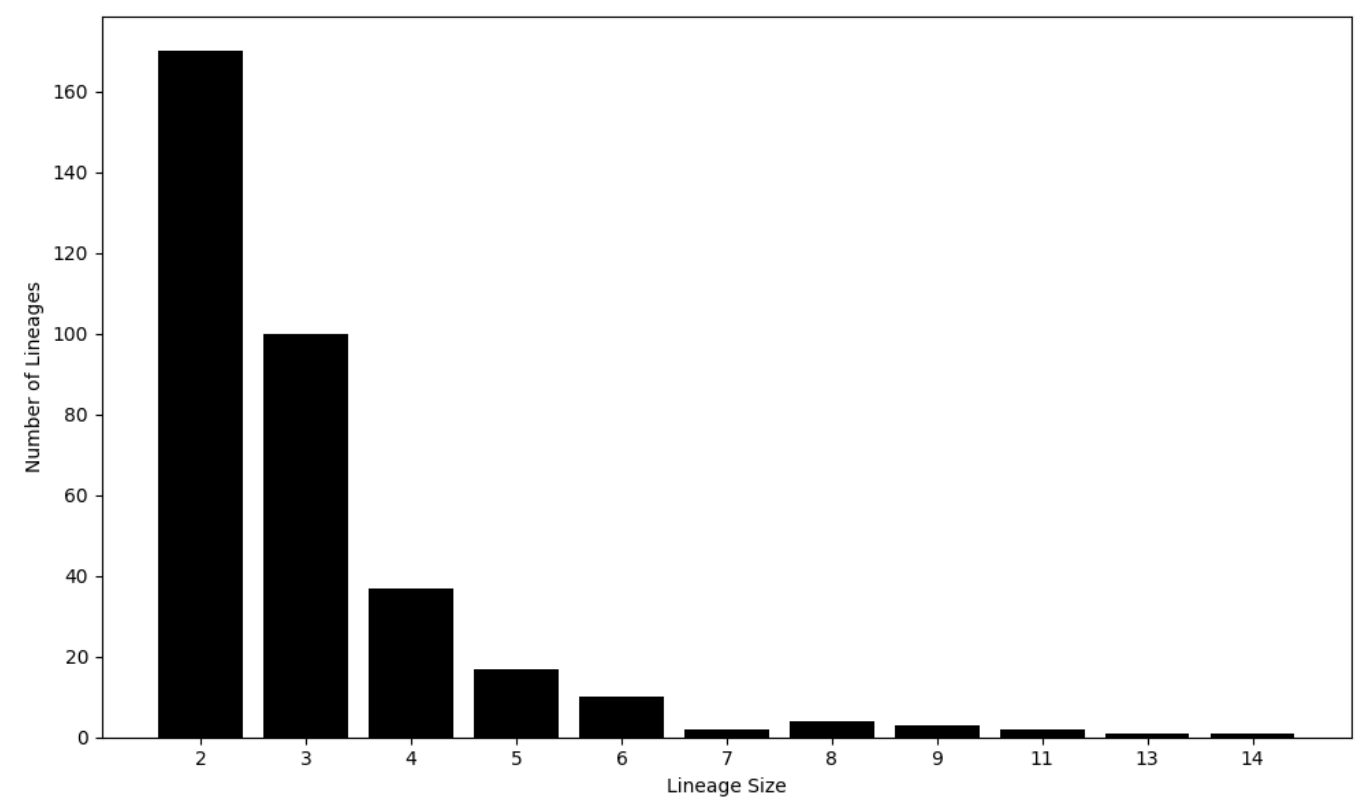

Fig. 2. Sizes of lineages in *lineageSet*

In figure 2, we classify lineages according to their size in terms of their number of contract versions. The lineage with the most number of contract versions encompasses 14 versions of that contract. one hundred seventy (170) smart contract lineages have only two (2) versions collected. It takes an average of 23 days to deploy a new version of a smart contract. We have restructured the lineages and the predecessor/successor relationships using open-source contracts. Additionally, we have paired the files and functions within these lineages. Approximately 88.73% of the Solidity files had at least one other version in the lineages. Additionally, we calculated the average similarity rate between these files, classified as predecessor-successor in 3450 pairs, and found an average rate of approximately 98%, whether comparing lines of code or entire files. In this file, we have 43964 pairs of contracts in which the data for the pie chart in Fig. 3 was derived from a comprehensive analysis of Solidity files, where each file was compared to its immediate predecessor to calculate the similarity rate. Specifically, Fig. 3 illustrates the distribution of similarity rates among pairs of Solidity files. Consequently, the pie chart categorizes the pairs into two distinct groups based on their similarity rates: those with a similarity rate of less than 90% and those with a similarity rate of 90% or higher.

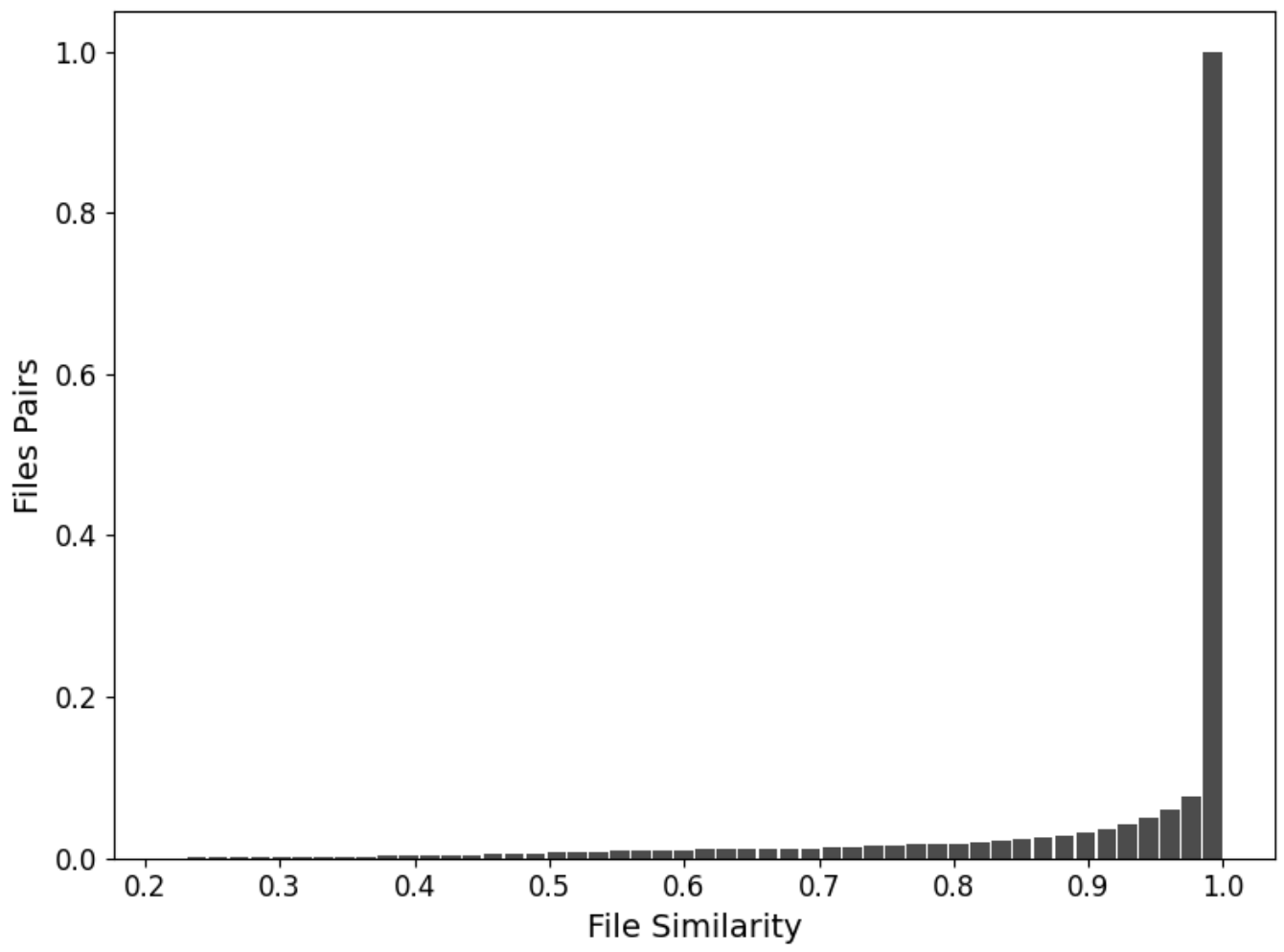

Fig. 3. Distribution of Similarity Rates in Predecessor/Successor Pairs of Solidity Files

The analysis reveals that a significant majority, 97%, of the pairs exhibit a similarity rate of 90% or higher. On the other hand, only 3% of the pairs have a similarity rate of less than 90%, suggesting a relatively small proportion of files with lower similarity. This rate quantifies the degree of similarity between two consecutive file versions, providing insight into the extent of changes made between versions. Additionally, we could further analyze the evolution of smart contract functionalities over time.

## VI. Demonstration: Building a dataset on Smart contract vulnerabilities and code changes

**Goal:** With this case study, we aim to conduct an empirical analysis of vulnerability management in smart contracts, leveraging smart contract lineages of *lineageSet*. To that end, we reuly on vulnerability detection tools that we apply on smart contract versions and track the appearance and disappearance of vulnerability warnings. Prior work has created a dataset by tracking GitHub smart contract projects that have vulnerability fix commits [13]. In contrast, we aim to construct a vulnerability lifecycle dataset based on deployed smart contracts available on Etherscan. Indeed, Etherscan ensures the authenticity of smart contracts by allowing users to verify the actual deployed code, while GitHub only provides access to the code without guaranteeing that it was deployed on the blockchain. Moreover, Etherscan offers more comprehensive access to smart contracts directly deployed on the Ethereum blockchain, ensuring complete data on contract interactions, including all transactions and event logs, which may not be fully captured on GitHub. Additionally, in terms of size, the GitHub-based approach is limited (46 projects). This motivates our study to focus on deployed contracts. We aim to adapt specific questions from the previous study regarding vulnerability distribution and first and last occurrences to the context of deployed smart contracts.

- *Q1.1* How many vulnerabilities are reported by the vulnerability analysis tools, and in how many Solidity files do they occur?
- *Q1.2* How many vulnerabilities have disappeared, and how many new vulnerabilities have been introduced?
- *Q1.3* How many vulnerabilities have been patched?

**Method:** Life cycle analysis requires smart contract versions, which are not readily available on Etherscan. We therefore rely on *lineageSet*. We employ analysis tools such as Slither [14], Mythril [15], and Conkas [16] to detect vulnerabilities in these versions. Slither and Mythril were chosen for their effective balance between performance and execution cost [17], and Conkas was integrated subsequently.
We analyzed this dataset to answer the questions.
**Results:**
*Vulnerabilities Distribution.* The dataset includes 79,677 data points detailing vulnerabilities across 384,676 vulnerable lines of code contained in 4,449 different files from 470 unique smart contracts (91.41%) distributed across 165 distinct lineages in *lineageSet*. This high rate of vulnerable smart contracts is obtained by combining the tools through a union operation and significantly decreases when the tool results are combined through the intersection. This finding is similar to the results of previous works that analyzed smart contracts with nine vulnerability detection tools [17], highlighting performance issues of vulnerability detection tools.
Figure 4 illustrates the distribution of distinct vulnerabilities, vulnerable files, and total detected vulnerabilities across various security analysis tools. A logarithmic scale is used to account for the significant variation in results. Each element in the diagram is represented by a specific color: blue for distinct vulnerabilities, gray for vulnerable files, and red for total vulnerabilities. The figure provides a visual breakdown of how vulnerabilities and vulnerable files are identified across the different tools.
The diagram shows the differences in the types and distribution of vulnerabilities detected, offering valuable insights into the effectiveness and coverage of each tool.
*Vulnerabilities life cycle.* 49.19% of the vulnerabilities represent newly introduced vulnerabilities, while 21.53% of them have disappeared in a successor version. 16.73% of updated files in open source contracts of *lineageSet*, we have at least a vulnerability disappearing without a newly introduced vulnerability. These files were distributed across 164 contract versions in 99 contract lineages. *Vulnerabilities persistence* We note an average of 283 days for vulnerabilities to disappear without a new introduction in lineage, while our previous finding underscores the need for Automated Program Repair (APR) tools.

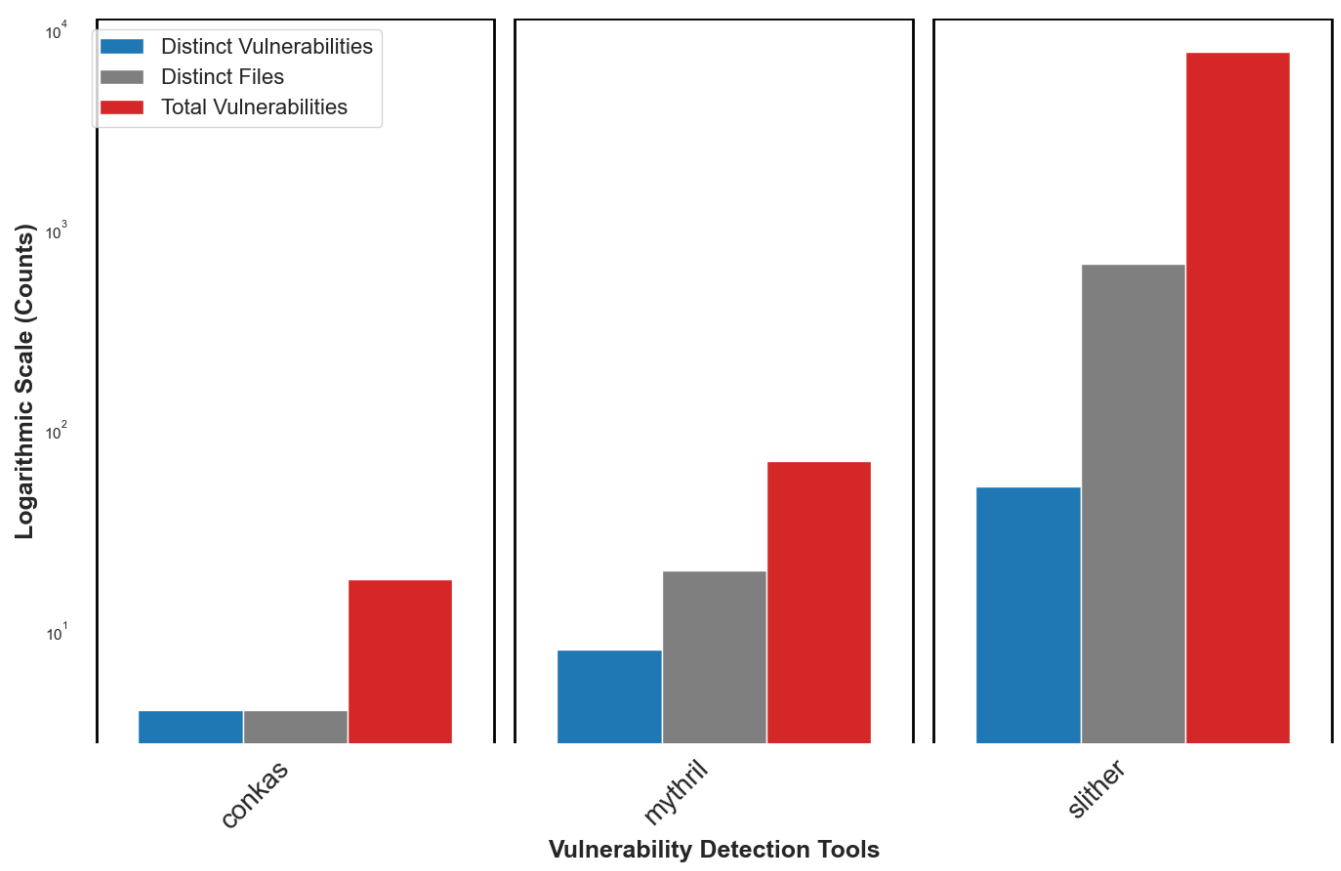

Fig. 4. Q1.1: Vulnerabilities and Vulnerable Files by Tool

## VII. Evaluation: Revisiting the Reliability of Similarity-Based Construction of Smart Contract Lineages

**Goal:** Our objective is to evaluate the effectiveness of similarity-based lineage construction methods used in prior research. To that end, we leverage *lineageSet*, built conservatively, as a ground truth dataset. We consider the locality-sensitive hashing (LSH) to be a measure of similarity. It is used in the Etherscan search engine.

**Methodology:** LSH is integrated within Etherscan, underlying its search engine, which allows for greater automation and flexibility in processing future Ethereum features. We compute smart contract similarity using Etherscan-based LSHimplementation. The similarity scores are expressed in three categories: Low, Medium, and High. Because our lineages in *lineageSet* are Ethereum smart contracts, they can serve as ground truth to discover false positives and false negatives based on the applied similarity thresholds of Etherscan. The experiment answers the following research question:

*RQ2: To what extent does the LSH-based approach accurately identify lineage relationships between Ethereum smart contracts?*

The LSH-based approach of the Etherscan search engine evaluates smart contract similarity by comparing their fingerprints. In practice, it is possible to make a request on Etherscan to obtain smart contracts that are similar to a given contract, with their degree of similarity categorized as low, medium, or high.

Given a smart contract $S$, the engine will output smart contracts that are similar. We consider them as candidates for being in the same lineage as $S$. To summarize:

(1) We collect contracts that are similar to $S$

(2) We define a similarity threshold $T$

(3) We form a lineage with $S$ and the collected contracts that have a similarity level to $S$ of at least $T$

**Results:** Since the Etherscan LSH-based similarity computation relies on fingerprints, we differentiate two types of smart contracts which may have different level of reliable fingerprints: open-source smart contracts tend to have more artifacts compared to non-open-source contracts, thus their fingerprints may be more accurate. The LSH-based approach defines various scenarios to form lineages. The differences between scenarios lie in the defined minimum similarity threshold $T$ that contracts in the same lineage must meet and whether the model targets all contracts or just open-source contracts. The evaluation methodology consists of three steps:

a) **Ground Truth Lineages Data**: *lineageSet* is used as the benchmark for evaluating the performance of the LSH-based method.

b) **Lineage Construction with the LSH Model**: The model is used to predict other contracts that belong to the same lineage as those in the Ground Truth Lineages Data (*lineageSet*) based on the aforementioned scenarios.

c) **Evaluation**: We compared the lineages formed by the model with the ground-truth lineages in *lineageSet*. The evaluation measured overall precision and recall for each scenario.

TABLE II
EVALUATION OF LSH-BASED APPROACH FOR SMART CONTRACT LINEAGE IDENTIFICATION

| Contract Type | Similarity threshold | Precision (%) | Recall (%) | Observations |
|---|---|---|---|---|
| Open-source | Low<br>Medium<br>High | 48.33<br>62.27<br>70.08 | 15.80<br>11.63<br>6.09 | Higher precision |
| All contracts | Low<br>Medium<br>High | 44.58<br>56.12<br>63.40 | 8.08<br>6.05<br>3.12 | Lower precision and recall |

Table II indicates a trade-off between precision and recall across different similarity thresholds, with open-source contracts showing better overall performance compared to all contracts combined. Indeed, open-source contracts exhibited higher precision across all similarity thresholds, starting at 48.33% at the low threshold and reaching 70.08% at the high threshold, although recall decreased significantly from 15.80% to 6.09%. In contrast, all contracts, including non-open-source ones, showed lower precision and recall. Their precision began at 44.58% at the low threshold and peaked at 63.40% at the high threshold, while recall declined from 8.08% to 3.12%. These findings indicate that focusing on open-source contracts improves precision, but recall presents a challenge as similarity thresholds increase. This trade-off underscores the importance of balancing precision and recall based on specific use cases and the availability of open-source data. This observation emphasizes the critical need to balance precision and recall when utilizing similarity computed by the LSH method for lineage formation.

> Key Insight
>
> **Low Recall Across All Scenarios:** Similarity-based approach to building lineages for smart contracts leads to low recall. When considering all contracts in Ethereum, the conservative proxy-based approach used to build *lineageSet* has a significantly higher precision and recall. Even when the required similarity is low, recall remains poor with LSH, including when considering open-source contracts.

The results of this approach align with previous research on code smart contract code reuse [3] which postulate that they have many code reuses. The findings of this case study also reinforce our relevance to construct lineages based on proxies.

## VIII. Discussion, Limitation and Related work

### A. Lineage construction

Our approach shares commonalities with those of Chen et al. [1] and Huang et al. [2].

TABLE III
COMPARISON OF OUR STUDY WITH PREVIOUS WORKS

| | SCLineage | Chen et al. [1] | Huang et al. [2] |
|---|---|---|---|
| **Targeted method of Contract upgrades** | Proxy-based upgrades | with self-destructed version | All |
| **Target deployed contracts** | Yes | Yes | Yes |
| **Automation** | Fully | No | Fully |
| **Granularity** | Contracts, files, and functions | Contract level | Contract level |
| **Similarity-based** | No | Yes | Yes |
| **Corpus availability** | All contract addresses | Some (self-destructed) | Any |
| **Continuously updated** | Yes | No | No |

A key advantage of our approach lies in its automation. This ensures scalability and efficiency and reduces the risk of human error. Additionally, the implementation details and results of our work are publicly available on GitHub. Despite the technical differences, we believe there is potential for synergy between our approach and those of Chen et al. and Huang et al. Their techniques could be complementary to ours, particularly for targeting updated contracts that do not utilize a proxy pattern. We plan to investigate the combination of these methods in future work.

### B. Lineage construction Implications

*lineageSet* extends previous empirical software engineering works on up-to-date repositories for smart contracts, such as Etherscan and Smart Corpus [4]. In java program, the Megadiff [18] study which is on code diff for Java programs enabled studies on automatic program repair (e.j [19]). *lineageSet* is similar to Megadiff and can therefore enable research in field of smart contract similar to those leveraging megadiff. In sum, *lineageSet* emerges as a large-scale, open-source, and reproducible dataset specifically designed for smart contract lineage analysis. The automation and public availability of our approach further enhance its usability and value for the research community.

However, it is important to acknowledge the limitations of our approach, which will be discussed in the following subsection.

## IX. Threats to validity

Our approach has inherent limitations, which we discuss and justify in the table below, explaining our design choices.

TABLE IV
THREATS TO THE VALIDITY OF `ContractTrace`.

| **Source** | **Risk** | **Rationale for Design Choice** |
|---|---|---|
| Proxy contracts only | Limited coverage | Ensures reliable upgrade tracking |
| Rule 3: Linear versioning assumption | Overlapping versions possible | Reduces complexity |

*Targeted contracts*: Our approach deliberately focuses on contracts upgraded through proxy mechanisms. This inevitably excludes lineages of contracts that were updated without proxies. Indeed, according to a study conducted in 2021 targeting 178 developers [8]: 39.39% of the selected respondents admitted that they discarded the old contract directly and deployed a new one, while 35.76% of them reported developing upgradable contracts. Thus, to be conservative, our dataset excludes many contract lineages. We did not employ similarity-based approaches, which could have allowed us to address various types of contracts. This is motivated by the homogeneity of Ethereum smart contracts, which is facilitated by code plagiarism, among other factors [3].

*Rule 3 Contract Version Ordering*: As a design choice, we opted for a linear versioning assumption within lineages. This implies that contract versions are assumed to be deployed chronologically, without overlapping activity periods. This decision was made with the objective of constructing a high-confidence dataset suitable for various applications. The linear model of this established rule allows us to confidently assert the relationships between the identified contracts. A more complex, tree-like structure would introduce intricate dependencies that could be challenging to verify. The weakness of this design choice is that we may exclude some versions in the lineages. In summary, our approach to collecting smart contract lineages introduces the potential to overlook certain contract versions within lineages. Additionally, we have excluded contracts that were not updated using the proxy method. However, these design choices are justified by our prioritization of a conservative approach that minimizes the inclusion of false positives and ensures the integrity of the lineage classifications.

## X. Conclusion

We addressed the challenge of identifying and classifying versions of smart contracts, which is crucial for various research areas in smart contract engineering. By applying a Design Science Research approach, we proposed `ContractTrace`, a novel infrastructure for systematically collecting and classifying smart contract lineages. It leverages the proxy pattern to minimize errors in lineage identification. Our up-to-date open-source dataset, *lineageSet* produced by `ContractTrace`, facilitates extensive research and analyses.

We demonstrated the utility of our approach in software engineering research, by conducting a case study focused on vulnerability tracking across smart contract versions, using *lineageSet*. Additionally, our methodology was evaluated through Locality-Sensitive Hashing (LSH) to test the effectiveness of clustering contract versions, confirming its robustness and relevance.

Building on the findings from both the case study and evaluation, we propose several directions for future research:

- `ContractTrace` and *lineageSet* provide valuable resources for advancing research in software engineering and security, facilitating more effective studies on smart contract vulnerabilities and their resolution. These tools can also be leveraged for developing smart contract repair techniques.
- Our findings from this case study reinforce those of previous works [13], [17] and emphasize the need for further exploration of Automated Program Repair (APR) techniques for smart contracts, particularly those with enhanced precision capabilities. When integrated with `ContractTrace`, these techniques could facilitate the creation of an up-to-date dataset that tracks code changes made to address vulnerabilities, similar to Big-Vul, a dataset focused on C/C++ code vulnerabilities [20].
- Investigating enhancements to the performance of LSH-based similarity metrics could improve the construction of contract lineages, extending their applicability to all types of contracts, irrespective of their update mechanisms. Future models can be tested using *lineageSet* as a ground-truth resource for lineage construction.

## XI. Acknowledgments

This work is supported by the Luxembourg Ministry of Foreign and European Affairs through their Digital4Development (D4D) portfolio under the project LuxWAyS (Luxembourg/West-Africa Lab for Higher Education Capacity Building in CyberSecurity and Emerging Topics in ICT4Dev.)

## References

[1] J. Chen, X. Xia, D. Lo, and J. Grundy, “Why do smart contracts self-destruct? investigating the selfdestruct function on ethereum,” *ACM Transactions on Software Engineering and Methodology (TOSEM)*, vol. 31, no. 2, pp. 1–37, 2021.

[2] Y. Huang, Q. Kong, N. Jia, X. Chen, and Z. Zheng, “Recommending differentiated code to support smart contract update,” in *2019 IEEE/ACM 27th International Conference on Program Comprehension (ICPC)*. IEEE, 2019, pp. 260–270.

[3] N. He, L. Wu, H. Wang, Y. Guo, and X. Jiang, “Characterizing code clones in the ethereum smart contract ecosystem,” in *Financial Cryptography and Data Security: 24th International Conference, FC 2020, Kota Kinabalu, Malaysia, February 10–14, 2020 Revised Selected Papers 24*. Springer, 2020, pp. 654–675.

[4] G. A. Pierro, R. Tonelli, and M. Marchesi, “An organized repository of ethereum smart contracts’ source codes and metrics,” *Future internet*, vol. 12, no. 11, p. 197, 2020.

[5] N. Szabo, “Formalizing and securing relationships on public networks,” *First monday*, 1997.

[6] S. Nakamoto, “Bitcoin: A peer-to-peer electronic cash system,” *Decentralized Business Review*, p. 21260, 2008.

[7] V. Buterin *et al.*, “A next-generation smart contract and decentralized application platform,” *white paper*, vol. 3, no. 37, pp. 2–1, 2014.

[8] J. Chen, X. Xia, D. Lo, J. Grundy, and X. Yang, “Maintenance-related concerns for post-deployed ethereum smart contract development: issues, techniques, and future challenges,” *Empirical Software Engineering*, vol. 26, no. 6, p. 117, 2021.

[9] Z. Gao, V. Jayasundara, L. Jiang, X. Xia, D. Lo, and J. Grundy, “Smartembed: A tool for clone and bug detection in smart contracts through structural code embedding,” in *2019 IEEE International Conference on Software Maintenance and Evolution (ICSME)*. IEEE, 2019, pp. 394–397.

[10] L. Jiang, G. Misherghi, Z. Su, and S. Glondu, “Deckard: Scalable and accurate tree-based detection of code clones,” in *29th International Conference on Software Engineering (ICSE’07)*. IEEE, 2007, pp. 96–105.

[11] C. K. Roy and J. R. Cordy, “Nicad: Accurate detection of near-miss intentional clones using flexible pretty-printing and code normalization,” in *2008 16th iEEE international conference on program comprehension*. IEEE, 2008, pp. 172–181.

[12] A. Gionis, P. Indyk, R. Motwani *et al.*, “Similarity search in high dimensions via hashing,” in *Vldb*, vol. 99, no. 6, 1999, pp. 518–529.

[13] Y. Wang, X. Chen, Y. Huang, H.-N. Zhu, J. Bian, and Z. Zheng, “An empirical study on real bug fixes from solidity smart contract projects,” *Journal of Systems and Software*, vol. 204, p. 111787, 2023.

[14] J. Feist, G. Grieco, and A. Groce, “Slither: a static analysis framework for smart contracts,” in *2019 IEEE/ACM 2nd International Workshop on Emerging Trends in Software Engineering for Blockchain (WETSEB)*. IEEE, 2019, pp. 8–15.

[15] *https://github.com/ConsenSys/mythril*.

[16] *https://github.com/nveloso/conkas*.

[17] T. Durieux, J. F. Ferreira, R. Abreu, and P. Cruz, “Empirical review of automated analysis tools on 47,587 ethereum smart contracts,” in *Proceedings of the ACM/IEEE 42nd International conference on software engineering*, 2020, pp. 530–541.

[18] M. Monperrus, M. Martinez, H. Ye, F. Madeiral, T. Durieux, and Z. Yu, “Megadiff: A dataset of 600k java source code changes categorized by diff size,” *arXiv preprint arXiv:2108.04631*, 2021.

[19] H. Ye, “Improving the precision of automatic program repair with machine learning,” Ph.D. dissertation, KTH Royal Institute of Technology, 2023.

[20] J. Fan, Y. Li, S. Wang, and T. N. Nguyen, “Ac/c++ code vulnerability dataset with code changes and cve summaries,” in *Proceedings of the 17th International Conference on Mining Software Repositories*, 2020, pp. 508–512.